\documentstyle[12pt]{article}

\catcode`\@=11 \@addtoreset{equation}{section}\catcode`\@=12

\begin{document}
\thispagestyle{empty}

\begin{center}
               RUSSIAN GRAVITATIONAL ASSOCIATION\\
               CENTER FOR SURFACE AND VACUUM RESEARCH\\
               DEPARTMENT OF FUNDAMENTAL INTERACTIONS AND METROLOGY\\
\end{center}
\vskip 4ex
\begin{flushright}                              RGA-CSVR-005/94\\
                                                gr-qc/9405067
\end{flushright}
\vskip 45mm

\begin{center}
{\bf
Multitemporal generalization of the Tangherlini solution }

\vskip 5mm
{\bf
Vladimir D. Ivashchuk and Vitaly N. Melnikov }\\
\vskip 5mm
     {\em Centre for Surface and Vacuum Research,\\
     8 Kravchenko str., Moscow, 117331, Russia}\\
     e-mail: mel@cvsi.uucp.free.msk.su\\
\vskip 60mm

           Moscow 1994
\end{center}
\pagebreak

\setcounter{page}{1}

\title{\vspace{-2cm}
Multitemporal generalization of the Tangherlini solution}
\author{Vladimir D. Ivashchuk$^{\dag}$ and Vitaly N. Melnikov $^{\dag}$}
\date{}
\maketitle
{\em Center for Surface and Vacuum Research, 8
Kravchenko str., Moscow, 117331, Russia}\\

The n-time generalization of the Tangherlini solution [1] is
considered. The equations of geodesics for the metric are integrated.
For $n = 2$ it is shown that the naked singularity is absent only
for two sets of parameters, corresponding to the trivial extensions of
the Tangherlini solution. The motion of a relativistic particle in the
multitemporal background is considered. This motion
is governed by the gravitational mass tensor. Some generalizations of
the solution, including the multitemporal analogue of
the Myers-Perry charged black hole solution, are obtained.


$^{\dag}$e-mail:mel@cvsi.uucp.free.msk.su
\pagebreak

\section{Introduction}

In ref. [3] the Tangherlini solution [1,2]  ($O(d+1)$-symmetric analogue
of the Schwarzschild solution) was generalized on
the case of $\bar{n}$ internal Ricci-flat spaces. The metric of this
solution is defined on the manifold
\begin{equation}
M = M^{(2+d)} \times M_{1} \times \ldots \times M_{\bar{n}},
\end{equation}
and has the following form
\begin{eqnarray}
g= &&- f^{a} dt \otimes dt + f^{b-1}  dR \otimes dR \\ \nonumber
   &&+ f^{b} R^{2} d \Omega^{2}_{d} +
       \sum_{i=1}^{\bar{n}} f^{a_{i}} g^{(i)},
\end{eqnarray}
where $M^{(2+d)}$  is $(2+d)$-dimensional space-time ($d \geq 2$),
$(M_{i},g^{(i)})$ is Ricci-flat manifold ($g^{(i)}$ is metric on
$M_{i}$), $dim M_{i} = N_{i}$, $i =1, \ldots , \bar{n}$;
$d \Omega^{2}_{d}$ is canonical metric on $d$-dimensional sphere
$S^{d}$,
\begin{equation}
f = {f}(R) = 1 - B R^{1-d},
\end{equation}
\begin{equation}
b = (1 - a - \sum_{i=1}^{\bar{n}} a_{i}N_{i})/(d-1),
\end{equation}
$B = const$, and the parameters $a, a_{1}, \ldots , a_{\bar{n}}$ satisfy
the relation
\begin{equation}
(a + \sum_{i=1}^{\bar{n}} a_{i}N_{i})^{2} +
(d-1) (a^{2} + \sum_{i=1}^{\bar{n}} a_{i}^{2} N_{i})= d.
\end{equation}
(Here the notations slightly differ from those of ref. [3]).
The metric (1.2) with the relations (1.3)-(1.5) imposed
satisfies the Einstein equations or, equivalently,
\begin{equation}
{R_{MN}}[g] = 0.
\end{equation}

Some special cases of the solution (1.2)-(1.5) were considered earlier
in the following publications: [4-6] ($d=2; \bar{n} =1; N_{1}=1$),
[7,8] ($d=2; \bar{n} = 2,3; N_{1}= \ldots = N_{\bar{n}}=1$),
[9] ($d=2; \bar{n} = 1$), [10] ( $\bar{n} = 1$; $a=
[(1- (d+N_{1})^{-1})/(1 - d^{-1})]^{1/2}$;
$a_{1} = -a/ (d+N_{1}-1)$); [11] ($ \bar{n} = 2$, $N_{2}=1$),
[12] ($ d = 2$; $\bar{n} $ is arbitrary).

It was shown in [3] that in the ($2+d$)-dimensional section of the
metric (1.2) a horizon exists only in the trivial case:
$a-1 = a_{1} = \ldots = a_{n}$ (this proposition was also suggested in
[12]). We also note that the cosmological analogue
of the solution [3] was presented in [14], where the tree-generalizations
of the solution were considered. (Such tree-generalizations may be also
constructed for spherically-symmetric case [3]).

In this paper we consider an interesting special case of  the
solution  [3].  This is the  $n$-time  generalization  of   the
Tangherlini solution. (The multitemporal analogue of the Schwarzschild
soluton ($ d = 2$) was considered earlier in [15].)

We  note,  that  the space-time manifolds  with extra
time dimensions were considered in gravitational context by
many authors  (see, for example, [17-26]).  Some  revival  of  the
interest in this direction was inspired recently by supergravity and
string models [22-27]. We note that the idea of the existence of
multidimensional domains with several times in the (multidimensional)
Universe was suggested by Sakharov in [20].

The paper is organized as following. In Sec. 2 the metric of the
multitemporal solution is considered. The explicit expression for the
Riemann tensor squared, corresponding to this solution is presented.
The proposition concerning the singularity of the solution at
$R=L > 0$ ($L$ is the parameter) for any set of dimensionless
parameters $(a_{i})$ except $n$ "tangherlinian" sets is
suggested. This proposition is proved for $n=2$. In Sec. 3
the equations of the geodesics, corresponding to the solution
are integrated. The notion of the multitemporal horizon is introduced.
It is proved that for all non-exceptional sets of $a_{i}$-parameters
the multitemporal horizon is absent. In Sec. 4 the motion of the
relativistic particle is considered. The multitemporal $O(d+1)$-
symmetric analogue of the Newton's formula for this case is obtained. In
multitemporal case the inertial and gravitational masses are
defined as matrices (or it may be defined also as tensors). In Sec. 5
the vacuum solution [3] is generalized on the electro-scalar-
vacuum case for the model with exponential scalar-electro-magnetic
coupling. Some infinite-dimensional generalizations of
the solution (including infinite-temporal and Grassmann-Banach analogues)
are presented.

\section{The metric}

We consider the special case of the solution (1.1)-(1.5) with
$\bar{n} = n - 1$  one-dimensional internal spaces:
$M_{i} =R$, $g^{(i)} = - dt^{i} \otimes dt^{i}$,
$i =1, \ldots , n-1$. Denoting $t =t^{n}$ and $a = a_{n}$ we get from (1.2)

\begin{eqnarray}
g=&& - \sum_{i=1}^{n} (1 - B R^{1-d})^{a_{i}} dt^{i} \otimes dt^{i} \\
  \nonumber
  && + (1 - B R^{1-d})^{b-1}  dR \otimes dR +
(1 - B R^{1-d})^{b} R^{2} d \Omega^{2}_{d},
\end{eqnarray}
where
\begin{equation}
b = (1 - \sum_{i=1}^{n} a_{i})/(d-1)
\end{equation}
and the parameters $a_{1}, \ldots , a_{n}$ satisfy the relations
\begin{equation}
(\sum_{i=1}^{n} a_{i})^{2} + (d-1) \sum_{i=1}^{n} a_{i}^{2} = d.
\end{equation}

The metric (2.1) with the parameters satisfying (2.2) and (2.3)
is the solution of the (n+d+1)-dimensional Einstein equations
(or,  equivalently, of the Ricci-flatness eqs. (1.6)).

The solution (2.1)-(2.3) is multitemporal ($n$-time) generalization
of the Tangherlini solution [1].

Let ${E}(d,n)$ be the set of points $a =  (a_{1}, \ldots , a_{n}) \in
R^{n}$, satisfying the relation (2.3). Clearly, that ${E}(d,n)$ is
ellipsoid. We denote the solution (2.1), corresponding to $a \in
{E}(d,n)$, $B \in R$ by $ g = {g}(a,B)$. An interesting fact is that the
metrics ${g}(a,B)$ and ${g}(-a,-B)$ are equivalent (for any $a \in
{E}(d,n)$ , $B \in R$), i.e.
\begin{equation} {g}(-a,-B)  = \varphi^{*}
{g}(a,B)
\end{equation} for some diffeomorphism  $\varphi$. This
diffeomorphism $\varphi = \varphi_{B}$ is defined by the relations
\begin{equation}
\varphi : (t^{i}, R_{*}, \theta^{\alpha}) \mapsto
(t^{i}, R, \theta^{\alpha}), \qquad R^{d-1} = R_{*}^{d-1} + B.
\end{equation}

Remark 1. An analogous equivalence takes place for the metric
(1.2).

Due to relation (2.4) it is quite sufficient to restrict our
consideration by the case $B > 0$ (the case $B = 0$ is trivial).

We introduce the following notations
\begin{equation}
T_{1} \equiv (1,0, \ldots , 0), \ldots , T_{n} \equiv (0, \ldots, 0,1),
\end{equation}
\begin{equation}
 T \equiv \{T_{1}, \ldots , T_{n}\} \subset {E}(d,n).
\end{equation}
We call the points (2.6) as Tangherlini points and the set (2.7)
as Tangherlini set. The metric (2.1) for $a = T_{k}$ has a rather
simple form
\begin{equation}
{g}(T_{k},B) = g^{(k)}_{T} - \sum_{i \neq k} dt^{i} \otimes dt^{i},
\end{equation}
where $g^{(k)}_{T}$ is the Tangherlini solution with the time
variable $t = t^{k}$, $k = 1,...,n$. The metric (2.8) is a trivial
(cylindrical) extension of the Tangherlini solution  with the time
$t^{k}$. It describes an extended membrane-like (string-like for
$n=2$) object. Any section of this object by hypersurface
$t^{i} = t^{i}_{0} =const$, $i \neq k$, is the $(2+d)$-dimensional
black hole [1,2], "living" in the time $t^{k}$.

{\bf Singularity}. The  Riemann tensor squared for the metric
(2.1) has the following form
\begin{eqnarray}
{I}[g] \equiv && R_{MNPQ}R^{MNPQ} = {\bar{I}}[g] /8 f^{2(b-1)}, \\
{\bar{I}}[g]  = && 16d(d-1) R^{-4} f^{-2} - 8d(d-1) R^{-2} f^{-1}
(b \frac{f'}{f} + \frac{2}{R})^{2}   \nonumber  \\
&& -d (b \frac{f'}{f} + \frac{2}{R})^{4} +
2d[2b \frac{f''}{f} - b(\frac{f'}{f})^{2} + \frac{2(b+1)}{R}
\frac{f'}{f}] ^{2}    \nonumber   \\
         && + \sum_{i=1}^{n} \{ -a_{i}^{4} (\frac{f'}{f})^{4}
+ 2a^{2}_{i} [ 2 \frac{f''}{f} + (a_{i} -1 -b)(\frac{f'}{f})^{2}]^{2}\}
 \nonumber   \\
 && +[d (b \frac{f'}{f} + \frac{2}{R})^{2} +
   \sum_{i=1}^{n} a_{i}^{2} (\frac{f'}{f})^{2}]^{2},
\end{eqnarray}
where $f' \equiv df/dR$ and $f$ is defined by (1.3).
We denote  $L = L_{B} \equiv  B^{1/(d-1)}$ for $B > 0$. The relation
(2.10) may be obtained from the formula presented in the
Appendix.

Proposition 1. Let $B > 0$ and $a = (a_{i}) \in {E}(d,n) \setminus T$, i.e.
the set of parameters $a$ is non-tangherlinian. Then the
quadratic invariant (2.9) for the metric (2.1) $g = {g}(a,B)$
is divergent: ${I}[g] \rightarrow \infty$, as $ R \rightarrow L$.

Proof. Here we prove the proposition for the case $n=2$. (The
case $n >2 $ will be considered in a separate publication [28]).

 From eqs. (2.9)-(2.10) we get the following asymptotical formula (here $n$
is arbitrary) \begin{equation} {I}[g] = \frac{A}{8} [{f'}(L)]^{4}
[{f}(R)]^{-2b -2} [1 + {O}(L-R)], \end{equation} as $R \rightarrow L$,
where
\begin{eqnarray} A = {A}(a) = && -db^{4} + 2d b^{2} + (db^{2} +
\sum_{i=1}^{n} a_{i}^{2})^{2} \nonumber \\
&& +\sum_{i=1}^{n} [-a_{i}^{4} + 2a^{2}_{i} (a_{i} -1 -b)^{2}].
\end{eqnarray}
The formula (2.11) is valid, when $A \neq 0$. We note  that
\begin{equation} (1 - r)/(d-1) \leq b \leq (1+r)/(d-1)  ,
\end{equation}
where $r \equiv \sqrt{dn/(d+n-1)}$
(see also Remark 2 below). It follows from (2.13) that
\begin{equation} 1 + b > 0
\end{equation}
and consequently (see (2.11))
${I}[g] \rightarrow \infty$  as $R \rightarrow L$, when
\begin{equation} A \neq 0
\end{equation}
Now, we prove the inequality (2.15) for $n= 2$ and
$a \in {E}(d,n) \setminus T$. For $n = 2$  we have
\begin{equation}
A = \frac{1}{2} d(d+1) b^{2} \bar{A},
\end{equation}
where
\begin{equation}
\bar{A} = - (d-1)(d+2) b^{2} + 8b  + 8.
\end{equation}
Using inequalities (2.13) it is not difficult to verify that
 $\bar{A} = {\bar{A}}(b) > 0$ (see also Remark 3 below).
On the other hand (in the case $n=2$) $b = {b}(a) = 0$ only
for the tangherlinian points $a = (0,1), (1,0)$. So, the inequality
(2.15) takes place for all $a \in {E}(d,2) \setminus T$ ($n=2$).
The proposition 1 is proved for $n = 2$.

Remark 2. In the coordinates
\begin{eqnarray}
&&\bar{a}_{1} = (a_{1} + \dots + a_{n})/\sqrt{n},  \nonumber \\
&&\bar{a}_{2} = (a_{1} - a_{2})/\sqrt{2},  \nonumber  \\
&&\bar{a}_{3} = (a_{1} + a_{2} - 2 a_{3})/\sqrt{6},
\qquad (n > 2)    \ldots   \nonumber    \\
&&\bar{a}_{n} = (a_{1} + \ldots + a_{n-1} - (n-1)a_{n})/ \sqrt{n(n-1)},
\nonumber
\end{eqnarray}
the ellipsoid equation (2.3) reads
\begin{equation}
(n +d -1) \bar{a}_{1}^{2} + (d-1) \sum_{i=2}^{n} \bar{a}_{i}^{2} = d.
\end{equation}
The inequalities (2.13) can be easily obtained from (2.18) and the
relation
\begin{equation}
b = (1 - \sqrt{n} \bar{a}_{1})/(d-1).
\end{equation}

Remark 3. The inequality $\bar{A} > 0$  may be proved,
using (2.13) and the following inequalities
\begin{eqnarray}
&& 1+ \sqrt{2d/(d+1)} < 5/2 < b_{+} (d-1),    \nonumber \\
&& 1 - \sqrt{2d/(d+1)} >- 1/2 > b_{-} (d-1),   \nonumber
\end{eqnarray}
where $b_{\pm} = [4 \pm \sqrt{8d(d+1)}]/ (d-1)(d+2)$ are zeros
of the quadratic polynomial ${\bar{A}}(b)$.

Thus for $n=2$, $a \in {E}(d,2) \setminus T$, $B > 0 $ the metric (2.1)
$g = {g}(a,B)$ is singular at $R = L$.

In the case $a \in T$, $B > 0 $ ($n$ is arbitrary) the metric (2.1)
$g = {g}(a,B)$ is regular for $R > 0$ and
\begin{equation}
{I}[g] = B^{2} R^{-2 -2d} d^{2}(d^{2} -1).
\end{equation}
Remark 4. In this case the metric has form (2.8). We remind that the
regularity of the Tangherlini metric for $R > 0$ may be easily seen using
the coordinates \begin{eqnarray} \bar{t} = t + \int dx
{\varphi}(x) ({f}(x))^{-1},
\qquad \bar{R} = R + \int dx ({\varphi}(x))^{-1} ({f}(x))^{-1},
\end{eqnarray}
where ${\varphi}(x) = (L/x)^{(d-1)/2}$.

\section{The geodesic equations}

We consider the geodesic  equations for the metric (2.1)
\begin{equation}
\ddot{x}^{M} + {\Gamma^{M}_{NP}}[g]\dot{x}^{N}\dot{x}^{P} = 0.
\end{equation}
Here and below $x^{M} = {x^{M}}(\tau)$ and $\dot{x}^{M}
= d x^{M}/d \tau $.

These equations are equivalent to the Lagrange
equations for the Lagrangian
\begin{eqnarray}
L &&= \frac{1}{2}{g_{MN}}(x)
\dot{x}^{M}\dot{x}^{N}  \nonumber  \\
  &&= \frac{1}{2} [ f^{b-1}(\dot{R})^{2}
  + f^{b} R^{2} {\kappa_{ij}}(\theta) \dot{\theta}^{i} \dot{\theta}^{j}
  - \sum_{i=1}^{n} f^{a_{i}}(\dot{t}^{i})^{2}].
\end{eqnarray} where the function $f = {f}(R)$ is defined in eq.
(1.3) and
\begin{equation}
\kappa = d\theta^{1} \otimes d\theta^{1}
+ \sin^{2}\theta^{1} d\theta^{2} \otimes d\theta^{2} + \ldots
+ \sin^{2}\theta^{1} \ldots  \sin^{2}\theta^{d-1}
   d \theta^{d} \otimes d\theta^{d}
\end{equation}
is standard metric on $S^{d}$. Here $ 0 < \theta^{1}, \ldots,
\theta^{d-1} < \pi$,  $0 < \theta^{d} = \varphi < 2 \pi$.

The complete set of integrals of motion for the Lagrange system (3.2)
is following
\begin{eqnarray}
&&f^{a_{i}} \dot{t}^{i} = \varepsilon^{i},  \\
&&f^{b}R^{2} \dot{\varphi} = j,
\end{eqnarray}
\begin{equation}
 f^{b-1}(\dot{R})^{2}  + j^{2} f^{-b} R^{-2}
  - \sum_{i=1}^{n} (\varepsilon^{i})^{2} f^{-a_{i}} = 2E_{L}
\end{equation}
$i = 1, ..., n$. We put here $\theta^{1} = \ldots =
\theta^{d-1} = \frac{\pi}{2}$. (This may be done for any
trajectory by a suitable choice of coordinate system.)
The radial equation
\begin{equation}
(f^{b-1}\dot{R})^{.}  + \frac{j^{2}}{2} (f^{-b} R^{-2})'
- \frac{1}{2} (\dot{R})^{2} (f^{b-1})'  -\frac{1}{2}
\sum_{i=1}^{n} (\varepsilon^{i})^{2} (f^{-a_{i}})' = 0
\end{equation}
(here $(.)' = d(.)/dR$) is generated by the Lagrangian
\begin{equation}
L_{R} = \frac{1}{2} [f^{b-1}(\dot{R})^{2}  - j^{2} f^{-b} R^{-2}
+ \sum_{i=1}^{n} (\varepsilon^{i})^{2} f^{-a_{i}}].
\end{equation}

We note, that the case $2E_{L} = 2L > 0$ in (3.6) correspond to a tachion.

{\bf Multitemporal horizon.} Here we  consider the null geodesics.
Putting $E_{L} =0$ in (3.6) we get for a light "moving" to the center
\begin{equation}
\dot{R} = - \sqrt{\sum_{i=1}^{n} (\varepsilon^{i})^{2} f^{1-b-a_{i}}
 - j^{2} f^{1-2b} R^{-2}}
\end{equation} and consequently
\begin{equation} t^{i} -
t_{0}^{i} = - \int_{R_{0}}^{R} dx \frac{\varepsilon^{i} [{f}(x)]^{-a_{i}}}
{\sqrt{\sum_{i=1}^{n} (\varepsilon^{i})^{2}
[{f}(x)]^{1-b-a_{i}} - j^{2} [{f}(x)]^{1-2b} x^{-2}}} ,
\end{equation}
$i = 1, \ldots ,n$.

Definition. Let $B > 0$, $\varepsilon = (\varepsilon^{i}) \neq 0$
and $a \in  E = {E}(d,n)$.  We say that the $\varepsilon$-horizon
takes place for the metric ${g}(a,B)$ at $R = L \equiv B^{1/(d-1)}$
if and only if
\begin{equation}
||t - t_{0}|| \equiv \sum_{i=1}^{n} |t^{i} - t_{0}^{i}|
\rightarrow + \infty,
\end{equation}
as $R \rightarrow L$ for all $t_{0}$ and $j$.

Proposition 2. Let $B > 0$,
and $a \in  E = {E}(d,n) \setminus T$. Then the $\varepsilon$-horizon
for the metric ${g}(a,B)$ at $R = L$ is absent for any
$\varepsilon \neq 0$.

Proof. We put $j =0$. It is sufficient to prove that all
integrals in  (3.10) are convergent, when $R \rightarrow L$ .
The integrals in (3.10) are convergent only if
\begin{equation}
s_{i} = - a_{i} - \frac{1}{2} min_{\varepsilon}(1-b -a_{i}) > -1
\end{equation}
for all $i \in K_{\varepsilon} \equiv \{j| \varepsilon^{j} \neq 0 \}$.
Here
\begin{equation}
min_{\varepsilon}(u_{i}) \equiv
min \{u_{i}| i \in K_{\varepsilon} \}.
\end{equation}
Indeed, the integrand in the $i$-th integral in (3.10)
behaves like $\varepsilon^{i} (L - x)^{s_{i}}$  as
$x \rightarrow L$. The set of inequilities (3.12) may be rewritten
as following
\begin{equation}
2a_{i} <  max_{\varepsilon}(a_{i}) + 1 + b,
\end{equation}
for all $i \in K_{\varepsilon}$, where $max_{\varepsilon}$ in defined
analogously to $min_{\varepsilon}$. It can be easily verified
that the set of inequalities (3.14) is equivalent to the  following
inequality
\begin{equation}
max_{\varepsilon}(a_{i}) < 1 + b.
\end{equation}
This inequlity follows from
\begin{equation}
   max(a_{i}) < 1 + b.
\end{equation}
Now we prove (3.16) for all $a \in  E \setminus T$. Let us
consider the tangent hypersurface to the ellipsoid $E$
in the point $T_{1} = (1,0, \ldots , 0)$. The equation
for this hypersurface has following form
\begin{equation}
d( a_{1}-1) + a_{2} + \ldots  + a_{n} = 0 .
\end{equation}
It is clear, that for all $a \in  E \setminus {T_{1}}$
\begin{equation}
d( a_{1}-1) + a_{2} + \ldots  + a_{n} < 0,
\end{equation}
or, equivalently,
\begin{equation}
   a_{1} < 1 + b.
\end{equation}
In analogous manner it may be proved that
\begin{equation}
  a_{i} < 1 + b.
\end{equation}
for all $a \in  E \setminus {T_{i}}$, $i = 1, ..., n$.
The inequalities (3.20) imply (3.16). The proposition is
proved.

Now we consider the case $a \in T$. Without loss of generality
we put $a = T_{1} = (1,0, \ldots , 0)$. It is not difficult to
verify  that in this case  the $\varepsilon$-horizon takes place
only if $\varepsilon^{1} \neq 0$.

\section{Relativistic particle}

Here we  consider the motion of the relativistic particle in
the gravitational  field, corresponding  to the  metric (2.1). The
Lagrangian of the particle is well-known
\begin{equation}
L_{1} = -m \sqrt{-{g_{MN}}(x) \dot{x}^{M} \dot{x}^{N}},
\end{equation}
where $m$ is the mass of the particle ($\dot{x}^{M}= dx^{M}/d \tau $).

The Lagrange equations for (4.1) in the proper time gauge
\begin{equation}
{g_{MN}}(x) \dot{x}^{M} \dot{x}^{N} = -1
\end{equation}
coincide with the geodesic equations (3.1). In this case
$(E^{i})= (m \varepsilon^{i})$ is the energy vector  and $J = mj$ is the
angular momentum (see (3.4) and (3.5)).  For fixed values of
$\varepsilon^{i}$ the (d+1)-dimensional part of the equations of motion is
generated by the Lagrangian
\begin{equation} L_{*} = \frac{m}{2}[f^{b}
 \bar{g}_{T, \alpha  \beta}(x) \dot{x}^{\alpha}\dot{x}^{\beta} +
\sum_{i=1}^{n} (\varepsilon^{i})^{2} f^{-a_{i}}], \end{equation} where
$\bar{g}_{T}$ is the space section of the Tangherlini metric.

 Now, we restrict our consideration by the non-relativistic
motion at large distances: $R \gg L_{B}$. In this approximation:
$t^{i} = \varepsilon^{i} \tau, \, \sum_{i=1}^{n}
(\varepsilon^{i})^{2} = 1.$
It follows from (4.3) that in this approximation we get a
non-relativistic particle of mass $m$, moving in the  potential
\begin{equation}
V = - \frac{m}{2} \sum_{i=1}^{n} (\varepsilon^{i})^{2}
\frac{a_{i}B}{R^{d -1}}
= - G \frac{m(\varepsilon^{i} M_{ij} \varepsilon^{j})}{R^{d-1}},
\end{equation}
where $G$ is the gravitational constant and
\begin{equation}
M_{ij} = a_{i} \delta_{ij} B/ 2G,
\end{equation}
are the components of the gravitational mass matrix.

It is interesting to note  that  the relation (4.4) may be
rewritten as following
\begin{equation}
V = -G \frac{ tr(M M_{I})}{R^{d-1}}
\end{equation}
where  $M_{I} = (m \varepsilon^{i} \varepsilon^{j})$ is the inertial
mass matrix of the particle.

The  solution (2.1) may be also rewritten in the matrix form
\begin{eqnarray}
g=&& -  [(1 - B R^{1-d})^{A}]_{ij} d \bar{t}^{i} \otimes d \bar{t}^{j}
\nonumber \\
&& + (1 - B R^{1-d})^{b-1}  dR \otimes dR +
(1 - B R^{1-d})^{b} R^{2} d \Omega^{2}_{d},
\end{eqnarray}
where $A$ is a real symmetric $n \times n$-matrix satisfying the
relation
\begin{equation}
(tr A)^{2}+ (d-1) tr (A^{2}) = d.
\end{equation}
and
\begin{equation}
b = (1 - tr A)/(d-1).
\end{equation}
Here $x^{A} \equiv  \exp (A \ln x)$ for $x > 0$. The metric (4.7) can
be reduced to the metric (2.1) by the diagonalization of the $A$-matrix:
$A= S^{T}(a_{i} \delta_{ij}) S$, $S^{T} S = 1_{n}$ and
the reparametrization of the time variables: $\bar{t}^{i} = S_{ij} t^{j}$.
In this case the gravitational mass matrix is
\begin{equation}
(M_{ij}) = (A_{ij} B/ 2G).
\end{equation}
We may also define the gravitational mass tensor as
\begin{equation}
{\cal M} = M_{ij} d \bar{t}^{i} \otimes d \bar{t}^{j}.
\end{equation}

We call the extended object, corresponding to the solution
(4.7)-(4.9) as multitemporal hedgehog. At large distances
$R^{d-1} \gg B$ this object
is described by the matrix analogue of the Newton's
potential
\begin{equation}
\Phi_{ij} = - \frac{1}{2} B R^{1-d} A_{ij} = -G R^{1-d} M_{ij}.
\end{equation}
Clearly that this potential for the diagonal case (2.1)
$A = a_{i} \delta_{ij}$ is a superposition of the potentials,
corresponding to "pure" black hole states (2.8).

Remark 5. It is interesting to note that the formula
\begin{equation}
A  = Q_{i} R^{1-d} dt^{i}
\end{equation}
describe the multitemporal ${O}(d+1)$-analogue of the well-known
electrostatic solution of the Maxwell equations. In this case the charge
$Q = (Q_{i})$ is a vector (or we may also define the charge as the  1-form
$Q_{i} dt^{i}$).

Remark 6. Let us consider the solution (2.1) for  $n=2$  with
$a_{1} >0$ and $a_{2} <0$. In  this case under a suitable
choice of the $\varepsilon^{i}$-parameters
a  point $R >L$,  may be a libration point, i. e. the point of
equilibrium. In this case
\begin{equation}
a_{1} (\varepsilon^{1})^{2}  +  a_{2} (\varepsilon^{2})^{2}
 [{f}(R)]^{a_{2}- a_{1}}  = 0  \nonumber
\end{equation}
and $\varepsilon^{2} \neq 0$.
An analogous situation takes place for arbitrary $n$, when
there exist positive and negative $a_{i}$-th parameters.

\section{Some generalizations}

Here we present some generalizations of the considered above
solutions. First, we consider the model described by the following
action

\begin{equation} S = \int d^{D}x \sqrt{|g|} \{
\frac{1}{2\kappa^{2}} {R}[g] -  \frac{1}{2\kappa^{2}} \partial_{M} \varphi
\partial_{N} \varphi g^{MN} - \frac{1}{4}\exp(2\lambda \varphi) F_{MN}
F^{MN}\},
\end{equation}
where $g = g_{MN} dx^{M} \otimes dx^{N}$ is the metric ,
$F = \frac{1}{2} F_{MN} dx^{M} \wedge dx^{N} = dA $ is the strength
of the electromagnetic field and  $\varphi$ is the scalar field.  Here
$\lambda$ is constant.  The action (5.1) describes for certain values of
parameters $\lambda$ and $D$ a lot of interesting physical models
including standard Kaluza-Klein theory, dimensionally reduced
Einstein-Maxwell theory, supergravity theories (see, for example [34]).
We present the spherically-${O}(d+1)$-symmetric solutions of the field
equations corresponding to the action (5.1) with the topology (1.1) [28].
The solution is following \begin{eqnarray} g= &&-
f_{1}^{(D-3)/{A}(\lambda)} f_{\varphi}^{2\lambda} dt \otimes dt
\nonumber \\
+ &&f^{-1/{A}(\lambda)}_{1} (f_{2}^{-1}
f_{\varphi}^{2\lambda} f^{2}) ^{1/(1-d)} [f_{2} du \otimes du +
d \Omega^{2}_{d}]   \nonumber  \\
+&& \sum_{i=1}^{\bar{n}}
f_{1}^{-1/{A}(\lambda)}  \exp(2 A_{i}u +2D_{i}) g^{(i)}, \end{eqnarray}
\begin{equation}
F =  Q f_{1} du \wedge dt,
\end{equation}
\begin{equation}
\exp \varphi = f_{1}^{(2-D)\lambda/ 2{A}(\lambda)} f_{\varphi}.
\end{equation}
In (5.2)-(5.4)
\begin{eqnarray}
&&f_{1} = {f_{1}}(u) = C_{1}(D-2)/ \kappa^{2} Q^{2} {A}(\lambda)
\sinh^{2}(\sqrt{C_{1}} (u-u_{1})) , \\
&&f_{2} = {f_{2}}(u) = C_{2}/(d-1)^{2} \sinh^{2}(\sqrt{C_{2}}(u-u_{2}))
,\\ &&f_{\varphi}= {f_{\varphi}}(u)= \exp (Bu +D_{\varphi}), \\
&&f = {f}(u) = \exp [ \sum_{i=1}^{\bar{n}} N_{i}( A_{i}u +D_{i})],\\
&&{A}(\lambda) = D-3 + \lambda^{2}(D-2),
\end{eqnarray}
and $Q \neq 0$, $D_{i}$, $D_{\varphi}$, $u_{1}$,
$u_{2}$ are constants and the parameters $C_{1}$, $C_{2}$, $B$, $A_{i}$
satisfy the the relation
\begin{eqnarray} \frac{C_{2}d}{d-1} =
&&\frac{C_{1}(D-2)}{D-3 + \lambda^{2}(D-2)} +B^{2} (1+\lambda^{2})
\nonumber   \\
&&+ \frac{1}{d-1} (\lambda B + \sum_{i=1}^{\bar{n}}
A_{i}N_{i})^{2} + \sum_{i=1}^{\bar{n}} A_{i}^{2} N_{i}.  \end{eqnarray}
The solution (5.2)-(5.10) generalizes the well-known Myers-Perry charged
black hole solution [2] for the model (5.1) on the case of $\bar{n}$
internal Ricci-flat spaces. (We remind that $(M_{i},g^{(i)})$ is
Ricci-flat space of dimension $N_{i}$, $i =1, \ldots , \bar{n}$.)
For $ \varphi = 0$ the presented here solution was obtained
in [29]  (the case  $d=2$ was considered previously in [30-32]).
Some special cases of this solution were considered also in [33,34].
For $\bar{n} = n - 1$,   $M_{i} =R$, $g^{(i)} = - dt^{i} \otimes dt^{i}$,
$i =1, \ldots , n-1$, $t =t^{n}$ we get from (5.2) the
multitemporal generalization of the solution [2] for the action (5.1).

In the zero charge case $F =0$ we have
\begin{eqnarray}
g= &&- \exp(2 A_{-1}u +2D_{-1}) dt \otimes dt
 \nonumber \\
+ && \exp[2(1-d)^{-1}\sum_{\nu} N_{\nu}( A_{\nu}u +D_{\nu})]
f_{2}^{1/(d-1)} [f_{2} du \otimes du + d \Omega^{2}_{d}]   \nonumber \\
+&& \sum_{i=1}^{\bar{n}}  \exp(2 A_{i}u +2D_{i}) g^{(i)},
\end{eqnarray}
\begin{equation}
\varphi = Bu +D_{\varphi}.
\end{equation}
The integration constants satisfy the relation
\begin{equation} \frac{C_{2}d}{d-1} =
 \frac{1}{d-1} ( \sum_{\nu} A_{\nu} N_{\nu})^{2} + \sum_{\nu}
A_{\nu}^{2} N_{\nu} + B^{2}.  \end{equation}
Hear $\nu = -1,1, \ldots , n $; $N_{-1} = 1$.

The solution (2.1)-(2.3) may be also generalized on the infinite-time
case: $n = \infty$. In this case the following restriction on the
parameters $a_{i}$ should be imposed (see also [14])
\begin{equation}
\sum_{i=1}^{\infty} |a_{i}| < +\infty.
\end{equation}
This relation implies
\begin{equation}
\sum_{i=1}^{\infty} |a_{i}|^{2} < +\infty.
\end{equation}
In this case the metric (2.1) is correctly defined on a proper
infinite-dimensional (Banach) manifold  and satisfies the Einstein
equations.  We note that an infinite-dimensional version of the Einstein
gravity was considered earlier by Kalitzin [16].

Remark 7. Another infinite-dimensional extension of the considered here
solution may be obtained if the field of real numbers $R$ is replaced
by the even part $G_{0}$ of the infinite-dimensional Grassmann-Banach
algebra $G =G_{0} + G_{1}$  [35,36]. In this case all coordinates and
the parameters of the solution (2.1) are elements of $G_{0}$.
(The $d$-dimensional sphere with the metric on it should be replaced by
its trivial $G_{0}$-extensions.)

\section{Conclusion}

In this paper the multitemporal analogue of the Tangherlini solution
was considered. It was shown that in the case of two time directions
the solution describes a naked singularity for any non-trivial
(non-tangherlinian) set of parameters. We have integrated the
geodesic equations for the considered solution. It was obtained
the multitemporal analogue for the Newton's formula ( eq. (4.6)),
describing the interaction between the massive particle and the
multitemporal extended object ("multitemporal hedgehog"), corresponding to
the solution.  It was shown that in the multitemporal case the inertial
and gravitational masses are matrices. (It may be defined also as
tensors).  We  have also obtained the  generalization of the Myers-Perry
charged black hole solution on the case  of a chain of
Ricci-flat internal spaces (this solution contains the  multitemporal
analogue as a special case).

\begin{center} {\bf Appendix}   \end{center}

Here  we present the expression for the tensor Riemann squared (2.9)
corresponding to the cosmological metric
\begin{eqnarray}
g= - {B}(t) dt \otimes dt +  \sum_{i=1}^{n} {A_{i}}(t) g^{(i)},
\nonumber
\end{eqnarray}
defined on the manifold $M = R \times M_{1} \times \ldots \times M_{n}$,
where $g^{(i)}$ is a metric on the manifold
$M_{i}$, $dim M_{i} = N_{i}$, $i =1, \ldots , n$. By a straightforward
calculation the following relation was obtained
\begin{eqnarray} {I}[g] =
&& \sum_{i=1}^{n} \{ A_{i}^{-2} {I}[g^{(i)}] + A_{i}^{-3} B^{-1}
\dot{A}_{i}^{2} {R}[g^{(i)}]  - \frac{1}{8}N_{i} B^{-2} A_{i}^{-4}
\dot{A}_{i}^{4}   \nonumber \\
&&+\frac{1}{4} N_{i} B^{-2}(2 A_{i}^{-1}
\ddot{A}_{i} - B^{-1} \dot{B} A_{i}^{-1} \dot{A}_{i} - A_{i}^{-2}
 \dot{A}_{i}^{2})^{2} \} + \frac{1}{8}  B^{-2} [\sum_{i=1}^{n} N_{i}
 (A_{i}^{-1} \dot{A}_{i})^{2}]^{2}.   \nonumber
\end{eqnarray}

\begin{center} {\bf Acknowledgments}   \end{center}

This work was supported in part by the Russian Ministry of Science.

\pagebreak

\end{document}